\documentclass{aa}

\usepackage{graphicx}
        
\begin{document}

\title{Hot C-rich white dwarfs: testing the DB--DQ transition
       through pulsations}

\author{A. H. C\'orsico\inst{1,2}\thanks{Member of the Carrera del 
       Investigador Cient\'{\i}fico y Tecnol\'ogico, CONICET, Argentina.},
       A. D. Romero\inst{1,2}\thanks{Fellow of CONICET, Argentina.}, 
       L. G. Althaus\inst{1,3}$^\star$, \and
       E. Garc\'\i a--Berro\inst{3,4}}
\institute{Facultad de Ciencias Astron\'omicas y Geof\'{\i}sicas,
           Universidad Nacional de La Plata,
           Paseo del  Bosque S/N,
           (1900) La Plata, Argentina    
           \and
           Instituto de Astrof\'{\i}sica La Plata, IALP, CONICET-UNLP
           \and
           Departament de F\'\i sica Aplicada,
           Escola Polit\`ecnica Superior de Castelldefels,
           Universitat Polit\`ecnica de Catalunya,
           Av. del Canal Ol\'\i mpic, s/n,
           08860 Castelldefels, Spain
           \and 
           Institut d'Estudis Espacials de Catalunya,
           c/Gran Capit\`a 2--4, 
           Edif. Nexus 104, 
           08034 Barcelona, Spain\\
\email{acorsico,aromero,althaus@fcaglp.unlp.edu.ar; garcia@fa.upc.edu}}

\offprints{A. H. C\'orsico}

\date{\today}

\abstract{Hot DQ white dwarfs  constitute a  new class of  white dwarf 
          stars, uncovered  recently within the framework  of the SDSS
          project. There exist nine of them, out of a total of several
          thousands   white   dwarfs   spectroscopically   identified.
          Recently, three  hot DQ white  dwarfs have been  reported to
          exhibit photometric variability with periods compatible with
          pulsation  $g$-modes.}   
         {Here, we  present a  nonadiabatic pulsation analysis  of the
          recently  discovered  carbon-rich hot DQ  white dwarf stars.  
          One  of  our main  aims  is  to  test the  convective-mixing
          picture for  the origin of  hot DQs through  the pulsational
          properties.}
         {Our study relies  on the full evolutionary models  of hot DQ
          white dwarfs  recently developed by Althaus  et al.  (2009),
          that  consistently  cover   the  whole  evolution  from  the
          born-again   stage  to  the   white  dwarf   cooling  track.
          Specifically, we present a stability analysis of white dwarf
          models  from  stages  before   the  blue  edge  of  the  DBV
          instability  strip ($T_{\rm eff}  \approx 30\,000$  K) until
          the domain of the hot DQ white dwarfs ($18\,000-24\,000$ K),
          including the transition  DB$\rightarrow$hot DQ white dwarf.
          We explore evolutionary  models with $M_*= 0.585\, M_{\odot}$
          and $M_*=  0.87\,M_{\odot}$, and two values  of the thickness
          of  the He-rich envelope  ($M_{\rm He}=  2\times 10^{-7}M_*$
          and $M_{\rm  He}= 10^{-8}  M_*$). These envelopes  are $4-5$
          orders of magnitude thinner  than those of standard DB white
          dwarf  models  resulting  from canonical  stellar  evolution
          computations.}
         {We found that at evolutionary phases in which the models are
          characterized  by  He-dominated  atmospheres,  they  exhibit
          unstable $g$-mode  pulsations typical of  DBV stars, and when the
          models   become  DQ   white  dwarfs   with  carbon-dominated
          atmospheres, they continue being pulsationally unstable with
          similar  characteristics than  DB models,  and  in agreement
          with the  periods detected in variable hot  DQ white dwarfs.
          In  particular, for  models with  $M_{\rm He}=  10^{-8} M_*$
          there  exists  a  narrow   gap  separating  the  DB  and  DQ
          instability domains.}
         {Our   calculations    provide   strong   support    to   the
          convective-mixing picture  for the formation of  hot DQs. In
          particular, our results point  to the existence of pulsating
          DB  white dwarfs  with  very thin  He-rich envelopes,  which
          after passing the DBV  instability strip become variable hot
          DQ stars.   The existence of  these DB stars with  very thin
          envelopes could be investigated through asteroseismology.}
  
\keywords{stars:   evolution   ---   stars:   interiors    ---  stars: 
         oscillations --- white dwarfs}

\authorrunning{C\'orsico et al.}

\titlerunning{Testing the DB--DQ white dwarf transition
              through pulsations}

\maketitle

 
\section{Introduction}
\label{intro}

The recent years  have witnessed a great deal of work  in the field of
white dwarf stars.  A recent  example is the unexpected discovery of a
population   of  white   dwarfs   characterized  by   carbon-dominated
atmospheres at effective temperatures between $\sim 18\,000$ and $\sim
24\,000$  K ---  also known  as  hot DQ  white dwarfs  (Dufour et  al.
2007).   This  finding immediately  attracted  the  attention of  many
researchers because the existence of  these stars poses a challenge to
the  theory of  stellar  evolution,  and could  be  indicative of  the
existence a new evolutionary channel of white-dwarf formation.

Dufour et al. (2008) proposed and evolutionary scenario to explain the
origin of hot DQ white dwarfs. In this scenario, undetected amounts of
He  ($M_{\rm He}\la  10^{-14}\,  M_{\odot}$) remaining  in  the C-  and
O-rich outer layers  of a PG1159 star like  the exotic object H1504+65
would  be  forced  to  float  at  the  surface  due  to  gravitational
separation,  leading  to a  He-dominated  white  dwarf,  first of  the
spectral  class  DO and  later  of the  spectral  class  DB.  In  this
picture, a C-rich atmosphere should eventually emerge as the result of
convective  mixing   at  lower  effective   temperatures.   The  first
quantitative  assessment  of  such {\sl  diffusive/convective  mixing}
scenario has been recently presented  by Althaus et al. (2009).  Using
full evolutionary  models that  consistently cover the  evolution from
the born-again stage  to the white dwarf cooling  track, these authors
presented strong theoretical evidence  supporting this picture for the
formation of hot DQs and the existence of an evolutionary link between
these stars  and the  PG1159 stars, including  H1504+65.  It  is worth
mentioning that the models of Althaus et al.  (2009) are characterized
by a  He content ranging  from $M_{\rm He}\sim 10^{-7}\,  M_{\odot}$ to
$10^{-8}\, M_{\odot}$.

Even  when the  impact of  the discovery  of DQ  white dwarfs  had not
declined,  Montgomery et  al. (2008)  reported on  the finding  of the
first variable  hot DQ star, SDSS J142625.70$+$575218.4  (with $\log g
\sim 9$  and $T_{\rm  eff} \sim 19\,800$  K), with a  confirmed period
$\Pi \approx 418$ s.  Shortly  after, Barlow et al. (2008) reported on
the  discovery   of  two  additional  variable  hot   DQ  stars,  SDSS
J220029.08$-$074121.5 ($\log g \sim  8$, $T_{\rm eff} \sim 21\,240$ K)
and  SDSS J234843.30$-$094245.3 ($\log  g \sim  8$, $T_{\rm  eff} \sim
21\,550$ K), with  periods $\Pi \approx 656$ s  and $\Pi \approx 1052$
s, respectively.  The measured  periodicities have been interpreted as
nonradial $g$-mode pulsations,  similar to the well-studied pulsations
of the GW Vir, V777 Her  and ZZ Ceti classes of white-dwarf variables.
The pulsation hypothesis, however,  has been defied by the possibility
that  these stars  could  be AM  CVn  systems, due  to the  similarity
exhibited  in the  pulse  shape  of the  light  curves (Montgomery  et
al.  2008).  On  the other  hand,  a compelling  argument against  the
interacting binary hypothesis is that  it does not explain why all hot
DQ white dwarfs are grouped within the same range of temperatures, and
none at higher  or lower effective temperatures (Dufour  et al. 2009).
Ultimately, the  confirmation of the pulsating nature  of the variable
hot   DQ   white   dwarfs   could   came   from   the   discovery   of
multiperiodicity.  Following  this  line  of  reasoning,  Fontaine  et
al.  (2009) have  recently announced  the discovery  of  an additional
period $\Pi \sim  319$ s (apart from the already known  at $418$ s) in
SDSS J142625.70$+$575218.4.

The theoretical quest of the origin of variability in hot DQ stars has
been addressed by Fontaine et  al.  (2008), who studied the hypothesis
that the  variability was due  to pulsations. Their  full nonadiabatic
analysis  revealed that  $g$-modes  can  be excited  in  the range  of
temperature where real DQs are  found (below $\sim 21\,500$ K) only if
the models are characterized by substantial amounts of He ($X_{\rm He}
\ga 0.25$) in their C-rich envelopes.  In a  subsequent effort, Dufour 
et al.   (2008) estimated the $T_{\rm  eff}$, $\log g$  and C/He ratio
for the  nine known hot DQ  stars and constructed  a dedicated stellar
model for  each object using the  same modeling as in  Fontaine et al.
(2008).   By using  a  nonadiabatic approach,  Dufour  et al.   (2008)
predicted that  only SDSS J1426$+$5752 should  exhibit pulsations, and
failed to  predict variability in SDSS  J220029.08$-$074121.5 and SDSS
J234843.30$-$094245.3.  However, it  appears that the pulsation models
of Fontaine et al.  (2008) and  Dufour et al.  (2008) are not entirely
consistent with their proposed  evolutionary picture for the formation
of  hot DQs.   Indeed, the  background models  they assumed  for their
stability  calculations  are characterized  by  a  He content  several
orders of  magnitude larger than the  content of He  required by their
evolutionary  scenario to  work (Dufour  et  al. 2008).  In fact,  the
bottom of the He-dominated envelope in their stellar models is located
at   a   fractional  mass   depth   of   $\log   q_{\rm  env}   \equiv
\log(1-M_r/M_*)= -3$ (Fontaine et al. 2008).

In this work, we perform  a nonadiabatic analysis of the pulsations of
hot  DQ white  dwarfs on  the basis  of the  full  evolutionary models
recently developed by Althaus et al.  (2009).  The models consistently
cover  the whole evolution  from the  born-again stage  to the  hot DQ
white dwarf domain.  As  it will be shown below, we are  able to get a
pulsational  picture of  the  variable  hot DQ  white  dwarfs that  is
entirely consistent with  the diffusive/convective mixing evolutionary
scenario  proposed for  their  formation. The  paper  is organized  as
follows. In  Sect. 2 we briefly  describe some aspects  of the stellar
evolution  and  pulsation modeling  we  employ  in  the present  work.
Sect.  \ref{escenario} is  devoted to  describe to  a some  extent the
diffusive/convective-mixing  scenario. We  elaborate on  our stability
analysis in Sect.   \ref{estabilidad}.  Finally, Sect. \ref{conclu} is
devoted to summarize our results.


\section{Stellar and pulsation modeling}
\label{modeling}

The stellar models employed in  this work have been generated with the
{\tt LPCODE} stellar evolutionary  code employed in our previous study
of  the  formation of  H-deficient  post-AGB  stars  via a  born-again
episode --- see Althaus et al.  (2005) and Miller Bertolami \& Althaus
(2006) for details  about the code. The code  is specifically designed
to compute the formation and  evolution of white dwarf stars.  In {\tt
LPCODE}, special emphasis is given  to the treatment of the changes of
the   chemical  abundances,   including  diffusive   overshooting  and
non-instantaneous  mixing,  which are  of  primary  importance in  the
calculation of  the thermal pulses  and born-again stage that  lead to
the formation of PG1159 stars.  Of relevance for the present study, we
mention that convection is treated  within the formalism of the mixing
length  theory as  given by  the ML2  parametrization (Tassoul  et al.
1990). We note  that this parametrization leads to  a theoretical blue
edge  of  the  DB   instability  strip  consistent  with  observations
(Beauchamp et al. 1999; C\'orsico et al. 2008).

The white dwarf evolutionary sequences employed in our stability study
correspond to those developed in  Althaus et al. (2009) to explore the
formation  of hot  DQs via  the diffusive/convective  mixing scenario.
The initial  stellar models for  those sequences are  realistic PG1159
stellar  configurations  derived  from  the full  evolution  of  their
progenitor stars  (Miller Bertolami \& Althaus  2006).  In particular,
we have considered sequences with  stellar masses of $0.87$ and $0.585
\,  M_{\odot}$.   The chemical  stratification  of  the initial  models
consists of a CO core, which is the result of core He burning in prior
stages, surrounded by  a He-, C- and O-rich  envelope, in agreement to
what is observed  in PG1159 stars. The fractional  mass of the He-rich
envelope  is in  the range  $10^{-8}\le  M_{\rm He}/M_*  \le 2  \times
10^{-7}$.  The  sequence with $M_*= 0.87\,  M_{\odot}$ was specifically
computed  by  Althaus  et  al.   (2009) to  explore  the  evolutionary
connection between H1504+65 and hot  DQ white dwarfs. For this stellar
mass,  the He content  of $2  \times 10^{-7}  M_*$ corresponds  to the
maximum He content expected in H1504+65 if we assume a post-born-again
origin  for  this  star  (Miller  Bertolami  \&  Althaus  2006).   The
evolutionary  calculations  have  been  computed  from  $T_{\rm  eff}=
100\,000$ K  down to the  domain of effective temperatures  typical of
hot DQs ($\sim 20\,000-17\,000$ K).

\begin{figure} 
\centering
\includegraphics[clip,width=250pt]{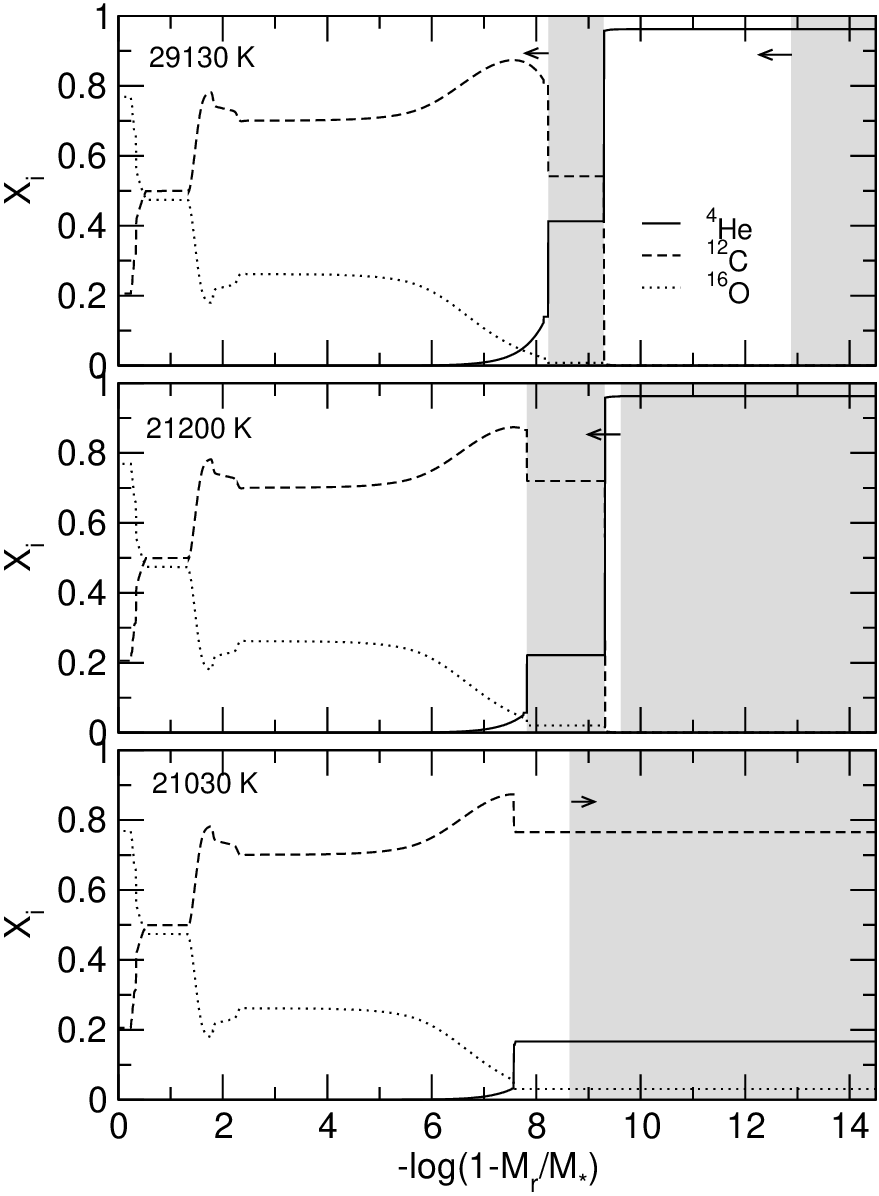}
\caption{The internal chemical abundances of helium, carbon and oxygen
         for models  with $M_*= 0.585  \, M_{\odot}$  and  $M_{\rm He}=
         10^{-8} M_*$ and effective temperatures  (from top to bottom) 
         of $\sim 29\,100, 21\,200$ and $21\,030$ K.  The grey regions 
         display convective zones.}
\label{figure1}
\end{figure}

The   pulsation   stability    analysis   was   performed   with   the
finite-difference nonadiabatic  pulsation code described  in C\'orsico
et al. (2006).  The code solves the full sixth-order complex system of
linearized equations and  boundary conditions as given by  Unno et al.
(1989).   Our  code  provides  the  dimensionless  complex  eigenvalue
($\omega$) and eigenfunctions according  the formulation given in Unno
et al.  (1989).  Nonadiabatic  pulsation periods and normalized growth
rates are evaluated as $\Pi = 2\pi/ \sigma$ and $\eta = -
\Im(\sigma)/  \Re(\sigma)$,  respectively.   Here,  $\Re(\sigma)$  and
$\Im(\sigma)$ are  the real and  the imaginary part,  respectively, of
the  complex eigenfrequency $\sigma  = (GM_*/R^3)^{1/2}  \omega$.  Our
code also computes the differential work function, $dW(r)/dr$, and the
running work  integral, $W(r)$,  as in Lee  \& Bradley  (1993).  These
functions are extremely useful for determining the driving and damping
regions  of the stellar  models.  As  in Fontaine  et al.   (2008) and
Dufour et al. (2008), our nonadiabatic computations rely on the 
frozen-convection approximation, in which  the perturbation of the convective
flux  is  neglected.   While  this  approximation  is  known  to  give
unrealistic  locations of  the red  edge of  instability, it  leads to
satisfactory predictions for  the location of the blue  edge of the ZZ
Ceti (DAV) instability  strip --- see Brassard \&  Fontaine (1999) ---
and also for the V777 Her (DBV) instability strip --- see Beauchamp et
al. (1999) and, more recently, C\'orsico et al. (2008).


\section{Evolutionary results: the diffusive/convective mixing scenario}
\label{escenario}

Let  us  briefly  summarize  the  diffusive/convective  mixing  picture
developed by  Althaus et al. (2009)  to explain the origin  of the hot
DQs.  In  this scenario,  the  starting  models  are realistic  PG1159
stellar configurations  with small contents of residual  He, being the
fractional masses in the range $10^{-8}\la M_{\rm He}/M_* \la 2 \times
10^{-7}$.  At the beginning of the hot phase of white-dwarf evolution,
the chemical abundance distribution of the envelope of PG1159 stars is
drastically modified by gravitational settling that forces He to float
at the surface  and heavier elements (C and O) to  sink.  As a result,
the star becomes  a DO white dwarf with a  very thin He-rich envelope.
With further cooling,  the star reaches the domain  of DB white dwarfs
($T_{\rm   eff}  \sim   30\,000$  K).    The  upper   panel   of  Fig.
\ref{figure1} shows  the internal abundance  distribution of $^{4}$He,
$^{12}$C,  and $^{16}$O  for  a  selected model  with  $M_*= 0.585  \,
M_{\odot}$, $M_{\rm He}= 10^{-8}  M_*$,  and $T_{\rm eff}= 29\, 130$ K.
Note  that,  at this  effective  temperature,  the  DB model  star  is
characterized  by  a inward-growing  outer  He  convection zone  (gray
region)  which is due  to the  recombination of   He{\sc ii}.   In addition,
there is a convective intershell  region at $8.2 \la -\log (1-M_r/M_*)
\la 9.3$. The existence of  this convection zone is the consequence of
the large opacity of the C-enriched layers below the pure He envelope.
As the  model star cools, the  bottom of this  convection region moves
inwards, thus increasing the C abundance there, as shown in the middle
panel of Fig.  \ref{figure1} that corresponds to a DB model at $T_{\rm
eff}= 21\,200$ K.   At this stage, the bottom  of the outer convection
region  is  very  close  to  the  upper  boundary  of  the  intershell
convective  region.  Shortly after,  both convection  zones eventually
merge,  and  a very  efficient  mixing  episode  --- favoured  by  the
presence of the  inthershell convection zone --- leads  to an envelope
rich in C and He ($X_{\rm C}= 0.77$, $X_{\rm He}= 0.17$), with a trace
abundance of  O ($X_{\rm O}=  0.03$). As a  result the DB  white dwarf
model  appears  as a  hot  DQ  white  dwarf.  The  chemical  abundance
distribution at  this stage  is depicted in  the bottom panel  of Fig.
\ref{figure1}  ($T_{\rm eff}=  21\,030$  K). 

The  evolution following  the merger of  the convection  zones is
  difficult to model. In fact, once the C-rich convective envelope has
  developed, it is  expected a substantial depletion of  carbon in the
  whole convective  envelope. Indeed,  the diffusion timescale  at the
  base  of  the  convection zone  of  the  newly  formed DQ  is  about
  $10^5-10^6$ yr, substantially shorter than the cooling timescale, of
  about $10^8$ yr.  Hence, we expect that the hot DQ stage is indeed a
  short-lived phase,  after which the star should  recover quickly its
  identity  as  a  He-atmosphere  DB  white  dwarf.   The  picture  is
  complicated  further  because  the  change  in  composition  in  the
  convective  envelope  will lead  to  changes  in  the depth  of  the
  convection  zone,  which will  affect,  in  turn,  the timescales  of
  diffusion. Our  numerical treatment of  diffusion does not  allow to
  properly follow  this feedback between convection  and diffusion. It
  is  not  discarded  that  the  star  experiences  additional  mixing
  episodes,  eventually  reaching   some  stationary  situation  below
  $T_{\rm eff} \sim 13\,000$ K in which the star would be a He-rich DB
  white dwarf  with traces of  C, that is,  a DQ white dwarf  like the
  ones observed at such effective temperatures (see, e.g., Bergeron et
  al.  2001). Additional  computations would  be needed  to  test this
  possibility.

\begin{figure*} 
\centering
\includegraphics[clip,width=450pt]{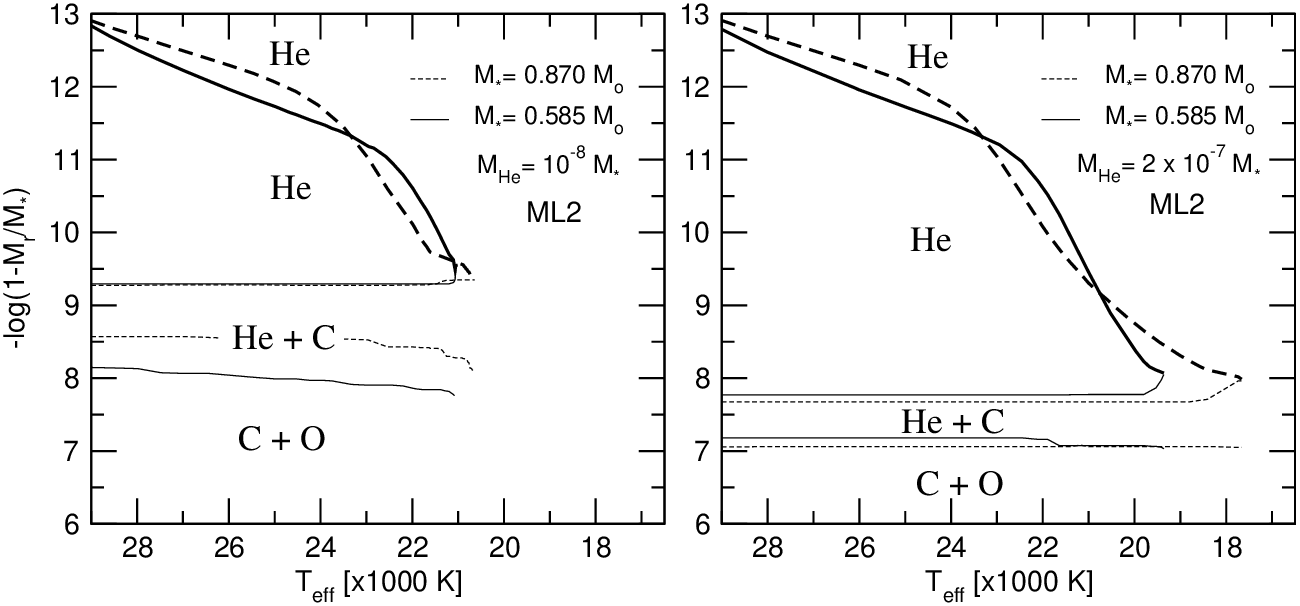}
\caption{The location  of the boundaries of the  convective regions in
  terms of  the effective temperature. Left  (right) panel corresponds
  to  the sequence  with $M_{\rm  He}=  10^{-8} M_*$  ($M_{\rm He}=  2
  \times  10^{-7}  M_*$).  Solid  (dashed)  lines  correspond  to  the
  sequence with $M_*= 0.585 \, M_{\odot}$ ($M_*= 0.870 \, M_{\odot}$).
  Thick lines are associated to the  location of the base of the outer
  convection zone, whereas      thin lines depict the  location of the
  boundaries of the convective carbon intershell. Stages after the merger
of the two convection zones are not shown.}
\label{figure2}
\end{figure*}

The  evolution  described  above  is  qualitatively  similar  for  the
remainder sequences analyzed  here. However, the effective temperature
at which the transition from DB to DQ white dwarf takes place, and the
final surface chemical abundances,  vary for different He contents and
different stellar  masses. For instance,  in the case of  the sequence
with $M_*= 0.585 \, M_{\odot}$ and  $M_{\rm He}= 2 \times 10^{-7} M_*$,
the transition  occurs at $T_{\rm  eff} \sim 19\,500$ K,  an effective
temperature somewhat lower  than for the case in  which the He content
is 20 times smaller ($T_{\rm  eff}\sim 21\,000$ K). On the other hand,
the final abundance of C at the surface is markedly smaller for larger
He  contents.  This  trend is  in  line with  observations (Dufour  et
al. 2008).   If we consider more  massive models and a  fixed value of
$M_{\rm He}$,  the transition  from DB to  DQ takes place  at somewhat
lower  effective temperatures  (Althaus  et al.   2009).  
To better clarify this point, we show in Fig. \ref{figure2} the
  evolution  of  the  convection  zones  in  terms  of  the  effective
  temperature for the  two stellar masses and He  contents analyzed in
  this paper.  Note that for effective temperatures below $23\,000$ K,
  the more  massive sequence is characterized by  deeper He convection
  envelopes.  But  for temperature less  than about $21\,000$  K, this
  trend is reversed and the  less massive sequence is characterized by
  a deeper He convection envelope.  By this effective temperature, the
  base of  the He convection  zone is close  to the outer edge  of the
  convective  intershell  region in  the  sequence  with $M_{\rm  He}=
  10^{-8} M_*$,  and so we expect that  the DQ is formed  at about the
  same effective temperature in both stellar masses.  For the sequence
  with $M_{\rm He}= 2 \times  10^{-7} M_*$, cleary the DQ formation is
  expected at a substantially  lower effective temperature in the more
  massive model.   Note also that the  rate of inward  increase of the
  base  of  the He  convection  zone by  the  time  it approaches  the
  convective  carbon intershell  is slowed  down. This  is due  to the
  presence  of  traces of  carbon  left  by  chemical diffusion.   The
  presence of trace carbon just above the outer edge of the convective
  intershell  region  leads  to  larger  radiative  opacities,  larger
  temperature gradients and  eventually to shallower convection zones,
  as compared with the behaviour expected in pure He envelopes.

Hitherto, we  have described  in some detail  the diffusive/convective
mixing scenario  that is able to  explain the formation of  the hot DQ
white dwarfs  from DB  white dwarfs with  thin He-rich  envelopes.  At
present,  observations firmly  indicate that  DB white  dwarfs exhibit
pulsations at effective temperatures between $23\,000$ K and $28\,000$
K  (the  DBV  instability  strip),  and  that  DQ  white  dwarfs  show
photometric variations  at $T_{\rm eff}  \approx 20\,000$ K.   At this
point, the question arises  to whether the diffusive/convective mixing
picture  may  also  be  able  to  predict  the  existence  of  the  DB
instability strip {\sl and} the DQ instability domain at the effective
temperatures  demanded by the  observations.  In  the next  section we
perform  a  stability  analysis  that  will help  us  to  answer  this
question.


\section{Stability calculations}
\label{estabilidad}

In this section we present  a detailed pulsation stability analysis of
our  set of DB  and DQ  white dwarf  models.  As  mentioned, numerical
difficulties prevented us from  following the further evolution of our
DQ  sequences far beyond  the transition  from DB  to DQ  white dwarf.
Thus, we are restricted to study  mainly the location of the blue edge
of the DQ instability domain, its dependence with stellar mass and the
content of He, and the  stability properties of DQ models located near
this blue  edge. In the  actual state of  affairs, the study  of these
important properties  characterizing the  variable DQ white  dwarfs on
the basis of a self-consistent evolutionary picture is very timely and
relevant.

We  analyze the  pulsational  stability of  about  800 stellar  models
covering a  wide range of effective temperatures  ($30\,000 \ga T_{\rm
eff} \ga 18\,000$ K) and stellar masses of $M_*= 0.585\, M_{\odot}$ and
$M_*= 0.870\, M_{\odot}$.  For each  mass value, we consider two values
of  the He  content: $M_{\rm  He}/M_*= 2  \times 10^{-7}$  and $M_{\rm
He}/M_*=  10^{-8}$.   For  each  stellar  model we  have  studied  the
stability of $\ell=  1$ $g$-modes with periods in  the range $50\ {\rm
s} \la  \Pi \la  3000$ s, thus  comfortably embracing the  full period
spectrum observed in variable DB and DQ stars.

\subsection{Template models}
\label{template}

\begin{figure}
\centering
\includegraphics[clip,width=250pt]{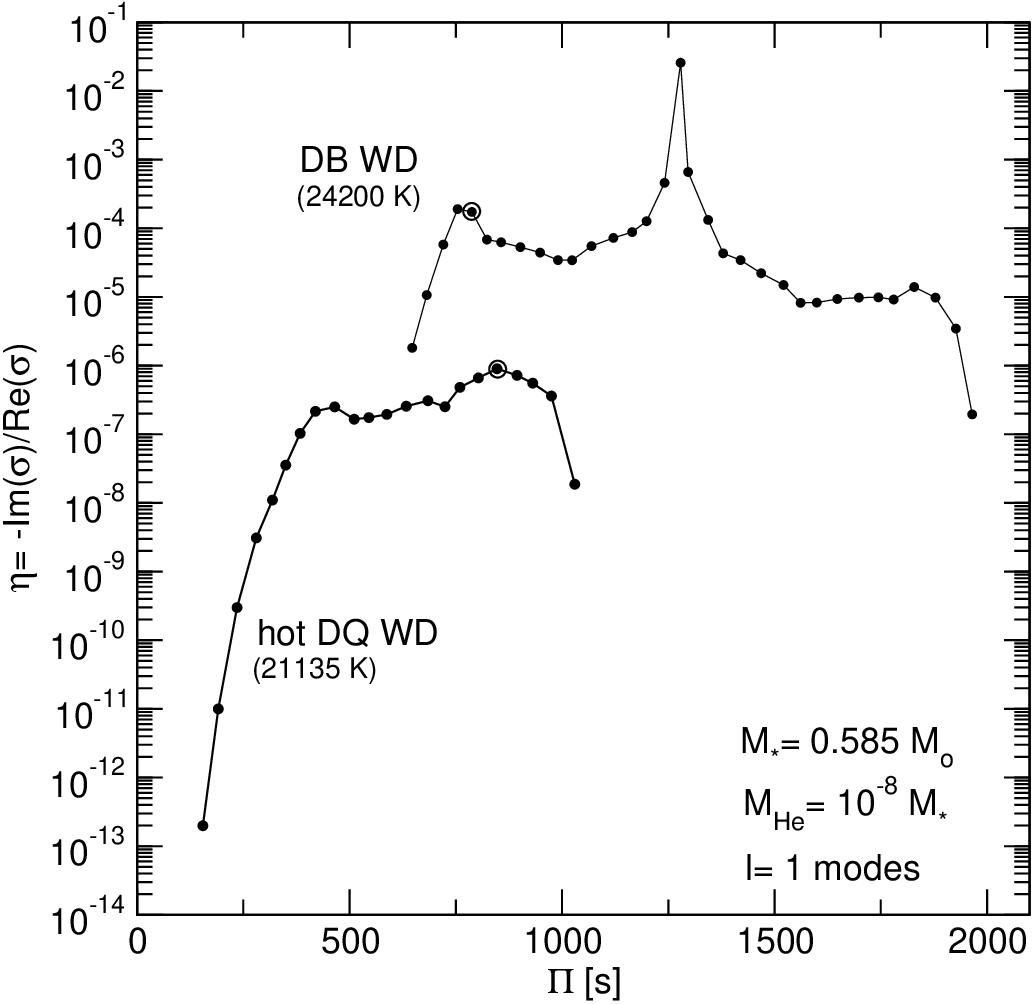}
\caption{The normalized  growth  rate in  terms of period (in seconds) 
         for overstable $g$-modes corresponding to a DB model ($T_{\rm
         eff}= 24\,200$ K) and a hot DQ white dwarf model ($T_{\rm eff}=
         21\,135$ K), both models corresponding to the same evolutionary
         sequence   ($M_*=  0.585 \,  M_{\odot}$   and   $M_{\rm   He}=
         10^{-8}M_*$).    For  each   model,  the   circumscribed  dot
         corresponds to the $k= 18$ $g$-mode.}
\label{figure3}
\end{figure}

We start by discussing the  stability properties of two template white
dwarf  models picked  out from  the evolutionary  sequence  with $M_*=
0.585 \, M_{\odot}$ and $M_{\rm He}/M_*= 10^{-8}$.  One of these models
is a  DB white dwarf model  ($T_{\rm eff}= 24\,200$ K),  and the other
one is  a hot  DQ white  dwarf model ($T_{\rm  eff}= 21\,135$  K). The
normalized growth rate ($\eta$)  in terms of pulsation periods ($\Pi$)
for overstable $\ell= 1$ modes  corresponding to these models is shown
in Fig.  \ref{figure3}.  Note that  modes excited in the DB model have
pulsation periods in the range $650 \la \Pi \la 2000$ s, substantially
longer than those excited in the  DQ model ($200 \la \Pi \la 1000$ s).
For the DB model, $\eta$ reaches a strong maximum at a 
period of about $1250$ s.  In average, for this  model the excitation  
is stronger for modes characterized  by short  periods, although at  
the edges  of the instability band the value of $\eta$ notably drops. 
On the other hand,
for the DQ  model the modes with longer periods  are more favoured for
excitation,  being the  value  of  the growth  rate  for the  shortest
periods more than seven order  of magnitude smaller than for the modes
with  longer  periods. We  note  that,  for  modes with  the  shortest
periods, the value of $\eta$ is so small ($\la 10^{-9}$) that they are
only marginally unstable.  Hence, hereinafter we shall consider that a
mode is unstable if $\eta \ga 10^{-9}$.

Clearly emphasized  in Fig.  \ref{figure3}  is the fact  that $g$-mode
excitation is  noticeably stronger  for the DB  model than for  the DQ
model.  Thus, the question arises about what would be the chance for a
given excited  mode in a  DQ star to  have time enough  for developing
observable amplitudes. In this respect,  we note that for our template
DQ model,  the maximum  and minimum $e$-folding  times for the  set of
excited modes are $\tau_e^{\rm max}=  459$ yr ($k= 4$, $\Pi= 280.5$ s,
$\eta= 3.1  \times 10^{-9}$)  and $\tau_e^{\rm min}=  5$ yr  ($k= 18$,
$\Pi=  847$  s,  $\eta= 9\times  10^{-7}$),  respectively\footnote{The
$e$-folding times are defined as $\tau_e \equiv 1/|\Im(\sigma)|$, such
that the time  dependence of the amplitude of  the pulsations is given
by  $\xi(t)  \propto  e^{i  \sigma  t}$, and  $\Im(\sigma)  <  0$  for
overstable  modes.}.   These  times   are  by  far  shorter  than  the
evolutionary timescale  at the DQ  stage, which ranges from  $10^7$ to
$10^8$  yr.  This  suggests that  the excited  modes in  hot  DQ white
dwarfs should have time enough  to reach observable amplitudes.  It is
worth mentioning that we arrive  at the same conclusion as Fontaine et
al. (2008),  who use a set  of DQ models and  a nonadiabatic pulsation
code completely independent from ours.

\begin{figure}
\centering
\includegraphics[clip,width=240pt]{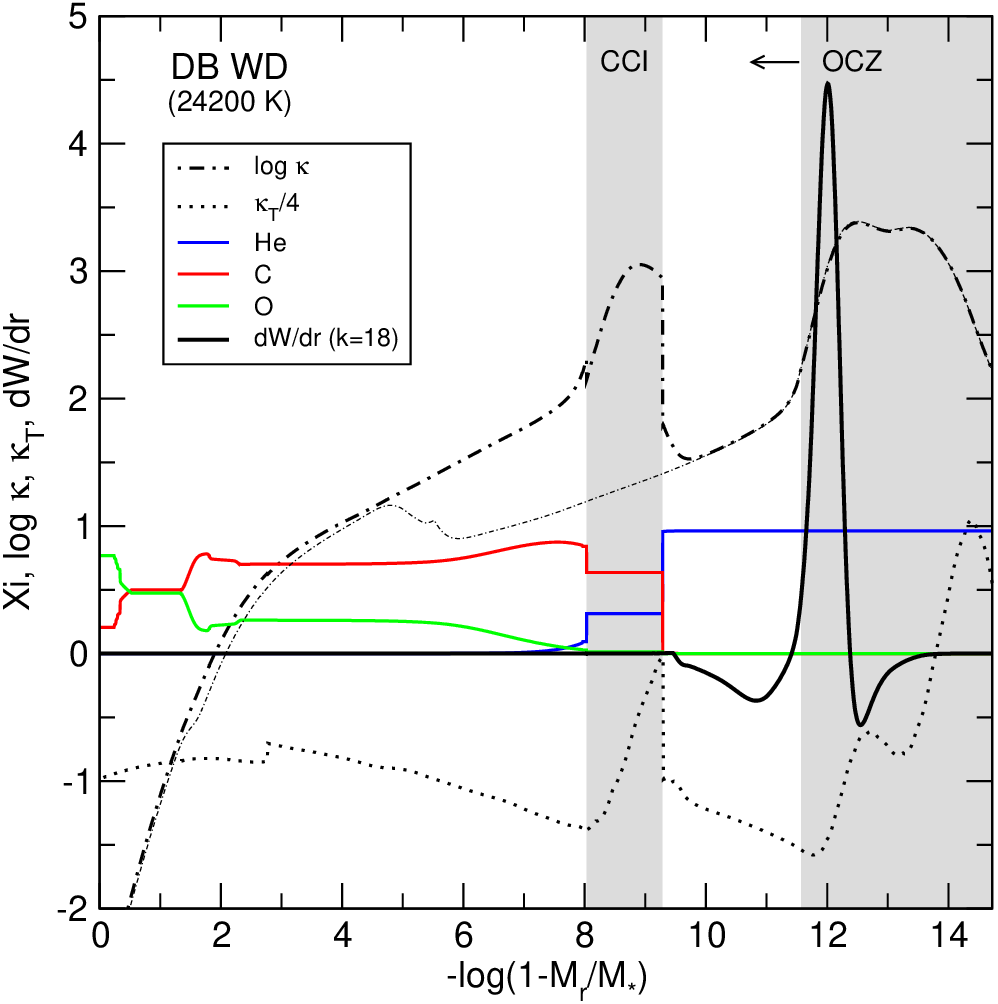}
\caption{The  logarithm of  the Rosseland opacity --- dot-dashed line 
         --- the opacity logarithmic derivative ($\kappa_{\rm T}$) ---
         dotted line --- and  the differential work function --- solid
         line --- for  a selected overstable $\ell= 1$,  $k= 18$ mode,
         in  terms  of  the  mass  coordinate,  corresponding  to  our
         $0.585\,   M_{\odot}$   DB    template   model   analyzed   in
         Fig. \ref{figure3}.  $dW/dr$ is in arbitrary units.  The gray
         regions  show the  locations  of the  convection zones.   The
         internal chemical abundances of helium, carbon and oxygen are
         also shown.}
\label{figure4}
\end{figure}

Fig. \ref{figure4} shows details of the driving/damping process in our
DB template model. The thick  solid line corresponds to $dW/dr$ for an
unstable  mode  with $k=  18$,  $\Pi= 787$  s  and  $\eta= 1.7  \times
10^{-4}$  (marked  as  a  circumscribed  dot in  the  upper  curve  of
Fig. \ref{figure3}).   Also plotted is the logarithm  of the Rosseland
opacity,  $\kappa$, and  its logarithmic  derivative,  $\kappa_{\rm T}
\equiv \left(\partial \log \kappa/\partial \log T \right)_{\rho}$.  As
can be seen, the region that destabilizes the mode (that where $dW/dr
> 0$) is located in a region slightly below the bump in the opacity at
$-\log(1-M_r/M_*) \sim 12.5$ ($T= 1.32 \times 10^5$ K), near the basis
of the  outer convection zone.   This bump corresponds to  the partial
ionization of He{\sc ii}. Above and below the driving region there are
two  zones of  damping.  Since the  contributions  to driving  largely
overcome the damping effects, the $k= 18$ mode is globally excited.

Note  also  the   presence  of  a  second  bump   in  the  opacity  at
$-\log(1-M_r/M_*) \sim  9$ ($T=  1.13 \times 10^6$  K), caused  by the
partial ionization of  C{\sc v} and C{\sc vi}.   Notably, this opacity
bump does not contribute at all  in the destabilization of the $k= 18$
mode.  So, in terms of pulsation stability, the mode does not ``feel''
the  presence  of the  C-bump.   For  illustrative  purposes, we  have
included in Fig. \ref{figure4}  the logarithm of the Rosseland opacity
(thin dot-dashed  line) corresponding to  a DB white dwarf  model with
the same  total mass  and effective temperature  than our  DB template
model, but characterized  by a thick He-rich envelope.  This model has
been extracted from  the calculations of C\'orsico et  al. (2008). The
run of the  opacity is coincident with that of  our DB template model,
except at  the location of  the C-bump, because the  thick-envelope DB
model does not have C in that region.
   
In summary, the  excitation of the overstable $k=  18$ $g$-mode of our
template  DB  model   is  not  affected  by  the   presence  of  C  at
$-\log(1-M_r/M_*) \sim 8-9$.  Since this property is shared by all the
overstable  $g$-modes of  the DB  models considered  in this  work, we
conclude that  overstable $g$-modes of  DB models having  thin He-rich
envelopes  are excited  in the  same way  as in  DB models  with thick
He-rich envelopes.   This property  suggests that the  DBV instability
strip could be  populated by DB white dwarfs with  both thin and thick
He-rich   envelopes.   This  possibility   could  be   tested  through
asteroseismological analysis performed on well-studied DBV stars.

\begin{figure}
\centering
\includegraphics[clip,width=240pt]{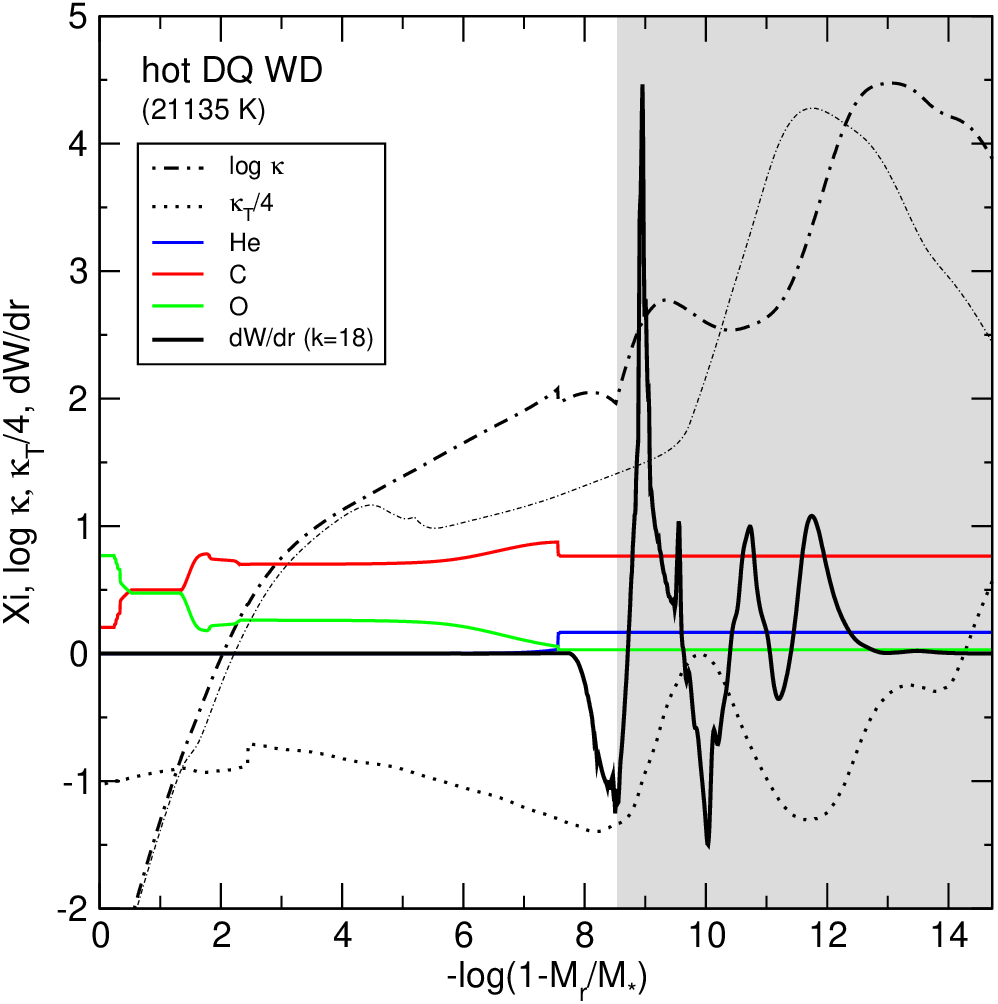}
\caption{Same as Fig. \ref{figure4}, but for our $0.585\, M_{\odot}$ DQ   
         template model analyzed in Fig. \ref{figure3}. The overstable
         mode has $\ell= 1$, $k= 18$ and $\Pi= 847$ s. }
\label{figure5}
\end{figure}

The  details of  the driving/damping  process in  our hot  DQ template
model for  a selected  overstable dipole mode  with $k= 18$  and $\Pi=
847$ s is displayed in Fig. \ref{figure5}.  This mode, which is marked
as a circumscribed  dot in the lower curve  of Fig. \ref{figure3}, has
the  largest growth rate  in our  DQ template  model ($\eta=  9 \times
10^{-7}$).  Note that, at variance with the DB template model, in this
case   most  of  excitation   comes  from   the  C-bump,   located  at
$-\log(1-M_r/M_*) \sim 9.5$  ($T= 1.177 \times 10^6$ K),  close to the
bottom of the outer convection zone.  We caution, however, that some 
of the driving at this region could be due to the neglect of the 
perturbation of the convective flux (the frozen-convection approximation),
as assumed in this work, instead of ``genuine'' driving due to 
the $\kappa$-mechanism acting at the C-bump. So, we would expect
the driving at this region to be somewhat reduced if the 
frozen-convection assumption were relaxed in the stability computations.
There are two additional driving
regions, which  also contribute to  the mode excitation,  
located at  $-\log(1-M_r/M_*) \sim  10.5$ and  $\sim 12$,
respectively. The most external one is  located at the hot side of the
opacity bump of the partial ionization of He{\sc ii}, corresponding to
$-\log(1-M_r/M_*) \sim 13$  and $T= 7.14 \times 10^4$  K. On the other
hand,  the presence of  the driving  region at  $-\log(1-M_r/M_*) \sim
10.5$ is not directly associated to any bump in the opacity. This is a
feature absent in  the models of Fontaine et  al. (2008).  Also, there
are three regions in our model  that contributes to the damping of the
mode,  instead of  two regions  as in  the models  of Fontaine  et al.
(2008). The total driving produced by the excitation regions overcomes
the damping effects, and the mode is globally unstable.

The characteristics of the driving/damping process described above for
a particular overstable  mode of our DQ template  model are also valid
for  all the  unstable  modes in  all  our hot  DQ  models.  Thus,  we
conclude that  overstable $g$-modes in  hot DQ white dwarf  models are
primarily  driven by the  strong destabilizing  effect of  the opacity
bump due to the partial ionization of  C{\sc v} and C{\sc vi} in a way
similar to what occurs in GW  Vir stars (C\'orsico et al. 2006), being
the  role  of  the partial  ionization  of  He{\sc  ii} of  much  less
relevance.   These  results  are  consistent with  those  reported  by
Fontaine  et  al.   (2008),  although  in  their  models  the  partial
ionization  of He{\sc  II}  plays a  more  relevant role  in the  mode
driving.

\subsection{The theoretical blue edges}
\label{theo}

\begin{figure} 
\centering
\includegraphics[clip,width=250pt]{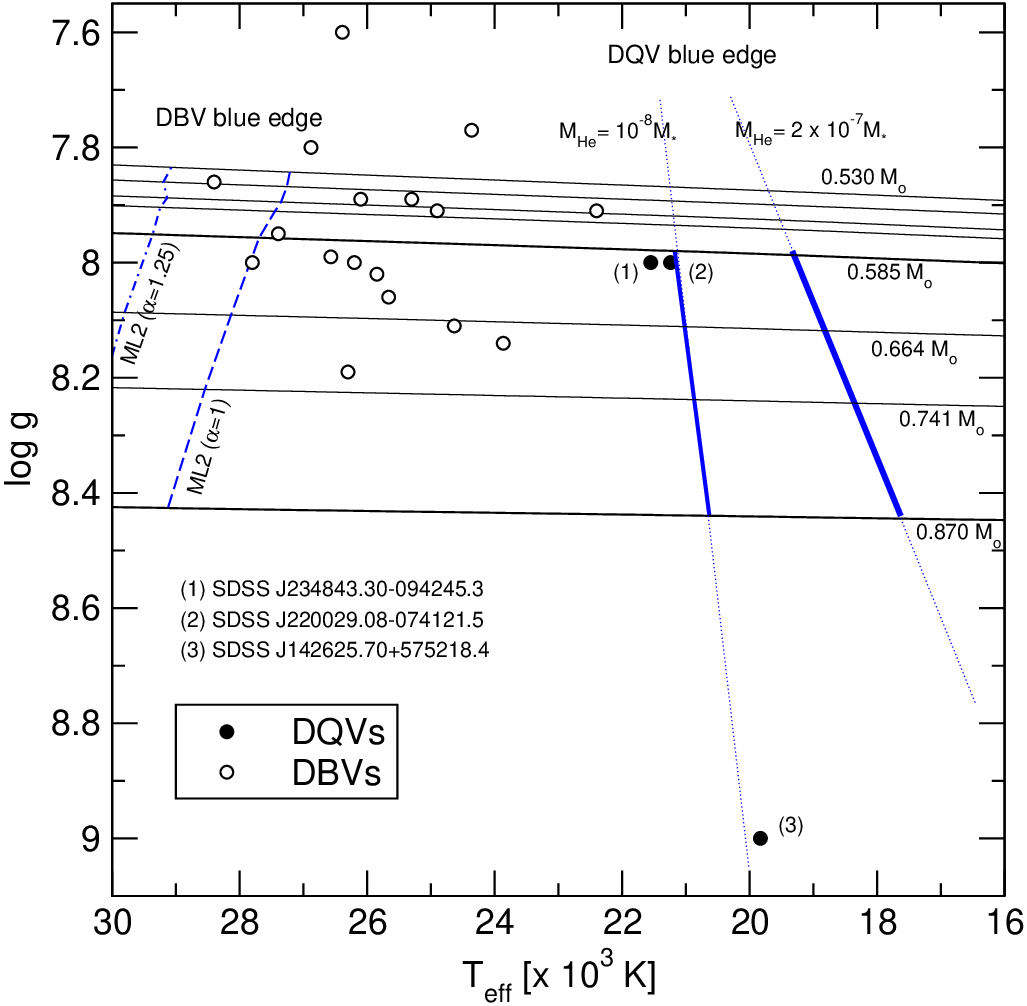}
\caption{A $\log g-T_{\rm eff}$ diagram  showing  the location  of the 
         recently  discovered three variable  DQs (the  ``DQVs'') with
         filled  circles,   and  the  known   seventeen  DBVs  (hollow
         circles).   Evolutionary  tracks  corresponding to  DB  white
         dwarf  models  with stellar  masses  of  0.530, 0.542, 0.556,
         0.565, 0.589, 0.664, 0.742  and $0.87\, M_{\odot}$ (from up to
         down) are  depicted with solid  curves. The blue edge  of the
         theoretical DBV instability strip --- taken from C\'orsico et
         al.  (2008)  --- is displayed with blue  dashed lines.  Also,
         the blue edge of the DQV instability domain (present work) is
         drawn  with  thick blue  solid  segments  for  two values  of
         $M_{\rm  He}$.  For  illustrative purposes,  these  lines are
         extended to high and low gravities (thin blue dotted lines).}
\label{figure6}
\end{figure}

Since  spectroscopic measurements  of white  dwarfs  provide effective
temperatures  and surface  gravities,  it  is useful  to  see how  the
evolutionary tracks  and the instability  domains look on  the $T_{\rm
eff} -  \log g$  plane.  In Fig.   \ref{figure6} we  plot a set  of DB
evolutionary  sequences on  this plane  (thin solid  lines), extracted
from the nonadiabatic study of  DBVs of C\'orsico et al. (2008). These
sequences correspond  to DB white dwarf models  characterized by thick
He-rich envelopes ($M_{\rm He}\sim 10^{-3}M_*$), as a result of which,
they never become DQ  white dwarfs.  Superimposed are two evolutionary
tracks ($M_*= 0.585\,  M_{\odot}$ and $0.870\, M_{\odot}$) corresponding
to  DB   white  dwarfs  with  thin  He-rich   envelopes  (thick  solid
curves). Note that, for a given  value of the stellar mass, the tracks
for  DB models with  thin and  thick He-rich  envelopes do  not differ
appreciably.

Fig. \ref{figure6}  includes the location of known  DBVs --- extracted
from Beauchamp et al. (1999) and  Nitta et al. (2009) --- and also the
known  three variable  DQ white  dwarfs (hereinafter  ``DQVs'').  Note
that  the observed blue  edge for  the DBV  instability strip  is well
accounted   for  by   a  theoretical   blue  edge   from  nonadiabatic
computations on DB models with the ML2 ($\alpha= 1$ or $\alpha= 1.25$)
flavor of the  Mixing Length Theory (MLT) of  convection (C\'orsico et
al. 2008).  Also  displayed in the figure is the  location of the blue
edge for the DQV instability  domain (thick solid segments). We recall
that our  DB and DQ  white dwarf models  assume the ML2  ($\alpha= 1$)
prescription  of the MLT.  We found  that our  DQ white  models become
pulsationally unstable  shortly after their formation, as  can be seen
by comparing the  blue edge of DQVs from  Fig.  \ref{figure6} with the
values of  effective temperature  at which DQ  white dwarf  models are
formed, see Table 1 of Althaus et al. (2009).

A  feature revealed by Fig. \ref{figure6} is that, in the
  current  computations, the  dependence  of the  blue  edge on  the
  stellar mass  of the DQVs  is the opposite  to that found  for the
  DBVs. That  is, we find that {\sl  the blue edge is  hotter for less
    massive DQV stars}.  This is due  to the fact that, in the context
  of  the  diffusive/convective mixing  scenario,  the temperature  at
  which the DQV blue edge appears is directly dependent on the stellar
  mass and  on the thickness of  the He-rich envelope of  the DB white
  dwarfs from  which the  DQ stars are  formed. In particular,  we have
  shown in Sect. \ref{escenario}  (see Fig. \ref{figure2}) that for
the He envelope masses assumed in our
  models, the merger of the two convective zones, and thus, the formation
  of the DQ white dwarfs, happens at higher effective temperatures for
  the less  massive models than for  the more massive  ones. Since the
  DQV  blue edge  occurs at  the $T_{\rm  eff}$ at  which the  DQs are
  formed, then this blue edge is hotter for less massive models.  Note
  that  this  trend is  at  odds with  the  results  of the  stability
  analysis of Fontaine et  al.  (2008) and also the order-of-magnitude
  estimation  obtained by Montgomery  et al.  (2008) from  the thermal
  timescale  ($\tau_{\rm th}$)  at the  base of  the  outer convection
  zone.

Fig.   \ref{figure6}   also  shows  that  the  blue   edge  is  hotter
($2\,000-3\,000$ K) for  DQVs with smaller contents of  He.  It is due
to the fact that the transition  DB to DQ white dwarf occurs early for
the case in which the He envelope  of the DB model is less massive, as
explained in Althaus  et al. (2009). Since the  pulsations are excited
immediately after the formation itself  of DQs, then the blue edge for
DQV stars is hotter for smaller contents of He at the envelope.

We close  this section by  noting that the  location of the  three DQV
stars in the $T_{\rm eff}-\log  g$ plane is qualitatively supported by
our  computations of  the  blue  edge for  the  case $M_{\rm  He}/M_*=
10^{-8}$. Actually,  only SDSS J142625.70$+$575218.4 lies  at right of
this theoretical blue edge. Note  that this star has a surface gravity
large   in  excess,   and   no  available   track   passes  near   its
location. Presumably, this star would  have a stellar mass $\sim 1.2\,
M_{\odot}$.  The  other two  DQVs, SDSS J220029.08$-$074121.5  and SDSS
J234843.30$-$094245.3, with a stellar  mass near $0.6\, M_{\odot}$, are
located  at  temperatures  slightly  higher than  the  predicted  blue
edge. Note, however,  that the blue edge could  be easily accommodated
at higher effective temperatures  by simply considering stellar models
with   somewhat  thinner  He-rich   envelopes  ($M_{\rm   He}/M_*  \la
10^{-8}$).

\subsection{The $T_{\rm eff}-\Pi$ plane}

\begin{figure} 
\centering
\includegraphics[clip,width=250pt]{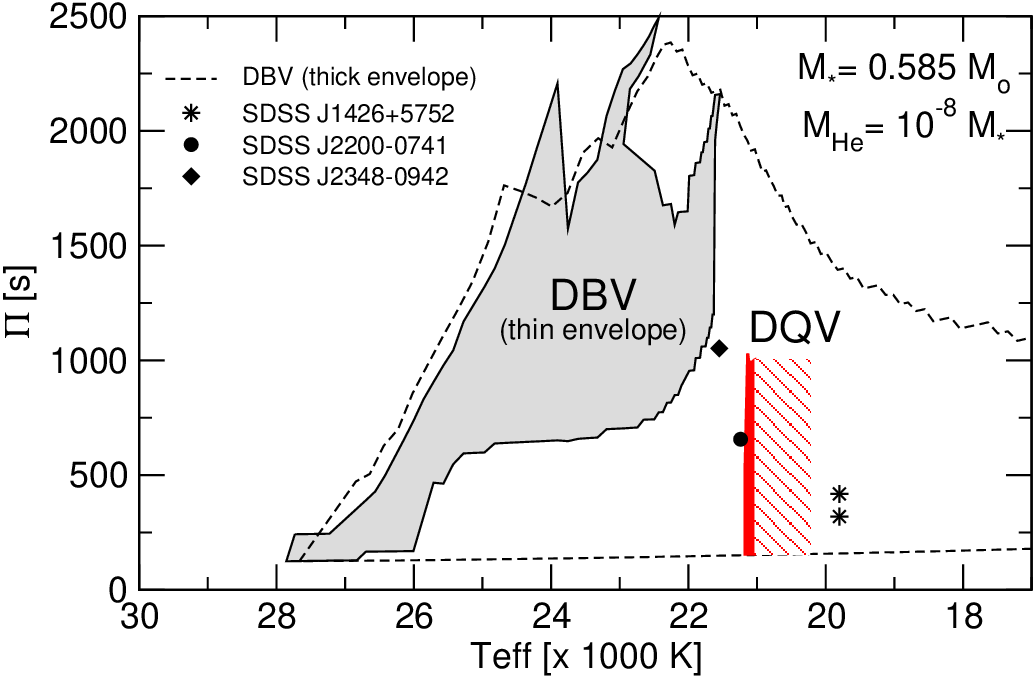}
\caption{The instability  domains on  the $T_{\rm eff}-\Pi$ plane  for  
         $\ell = 1$ $g$-modes corresponding  to our set of models with
         $M_*= 0.585 \, M_{\odot}$ and $M_{\rm He}= 10^{-8} M_*$.}
\label{figure7}
\end{figure}

\begin{figure}
\centering
\includegraphics[clip,width=250pt]{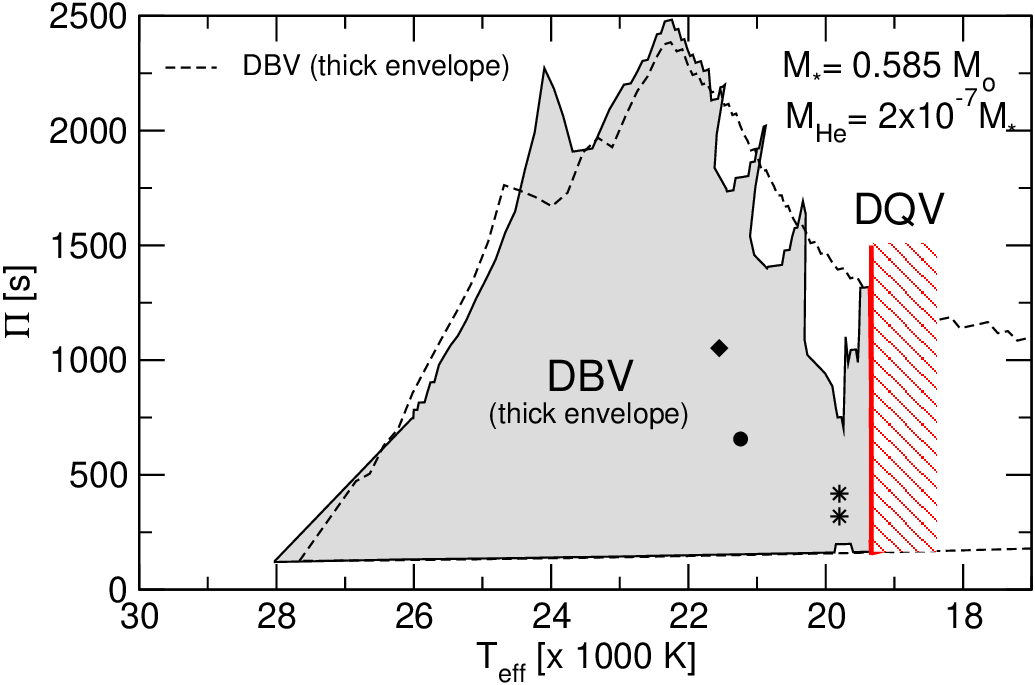}
\caption{Same as Fig. \ref{figure7}, but for models with  $M_{\rm He}= 
         2 \times 10^{-7} M_*$.}
\label{figure8}
\end{figure}

\begin{figure}
\centering
\includegraphics[clip,width=250pt]{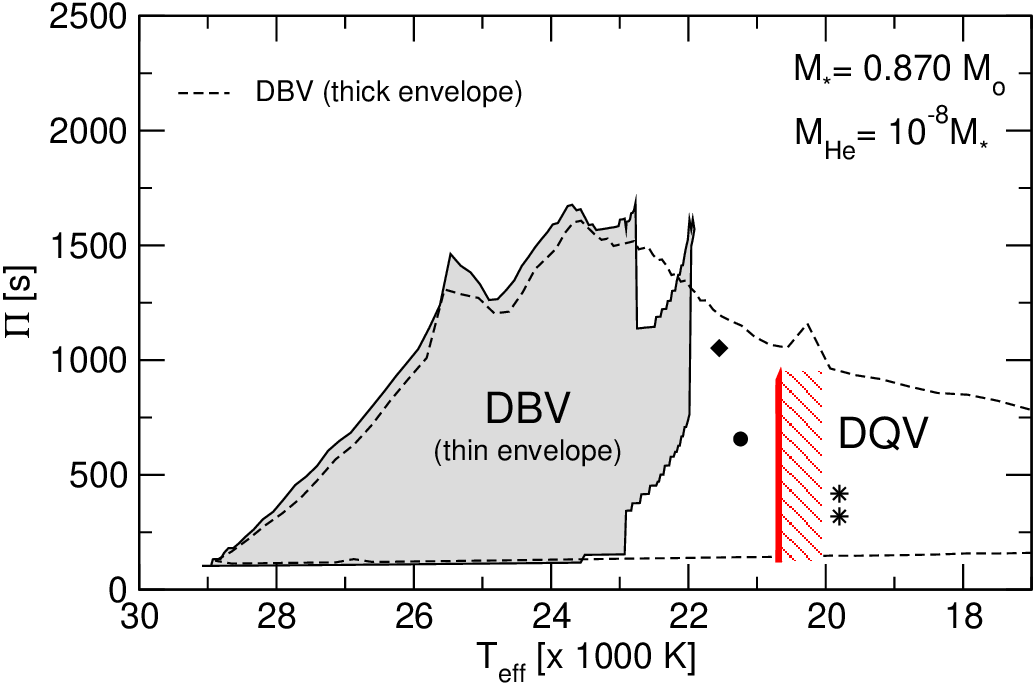}
\caption{Same as Fig. \ref{figure7}, but for models with $M_*= 0.870 \, 
         M_{\odot}$.}
\label{figure9}
\end{figure}

\begin{figure}
\centering
\includegraphics[clip,width=250pt]{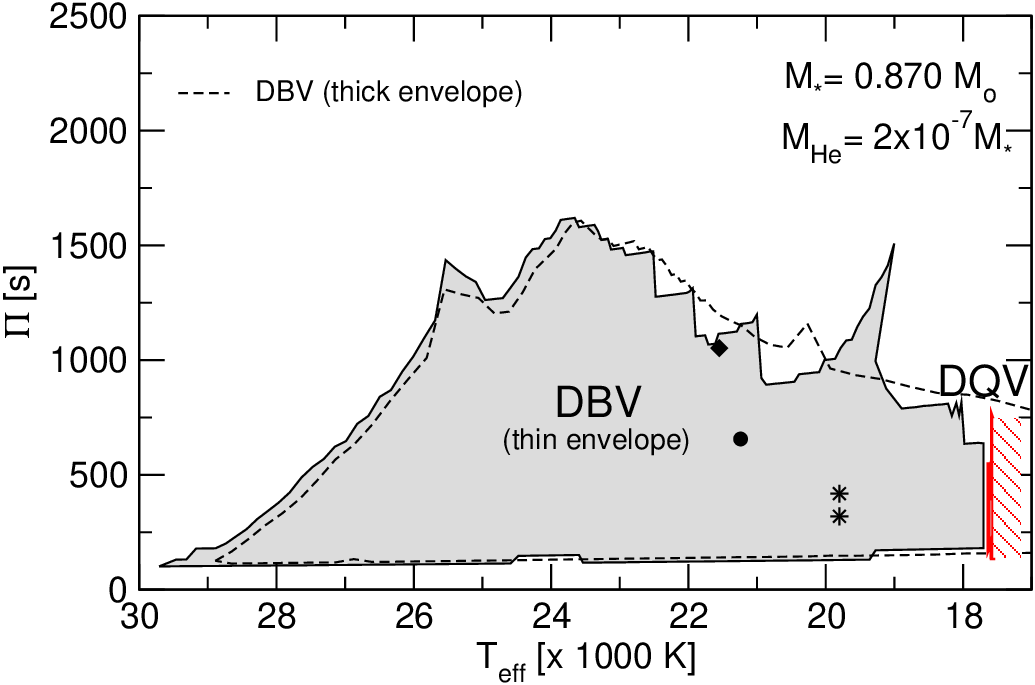}
\caption{Same as Fig. \ref{figure7}, but for models with $M_*= 0.870 \,  
         M_{\odot}$ and $M_{\rm He}= 2 \times 10^{-7} M_*$.}
\label{figure10}
\end{figure}

In this  section we  explore the domains  of unstable dipole  modes in
terms  of the  effective temperature.   Fig.  \ref{figure7}  shows the
instability domains on the $T_{\rm eff}-\Pi$ plane for the sequence of
$M_*=  0.585\,  M_{\odot}$ and  $M_{\rm  He}=  10^{-8}  M_*$.  The  DBV
instability domain is  shown as a grey area,  whereas the beginning of
the DQV  instability region is shown  as a vertical red strip. Also 
included in  the plot are
the periods detected in the three known DQV stars.  In the interest of
a comparison, we also include the instability domain of DB models with
thick  He-rich  envelopes ($M_{\rm  He}\sim  1  \times 10^{-3}  M_*$),
according     to      C\'orsico     et     al.      (2008)     (dashed
lines). Figs.  \ref{figure8}, \ref{figure9}, and  \ref{figure10} depict
the situation for the remainder  sequences considered in this work. In
each figure we have drawn a stripped region to illustrate the possible
extension to lower temperatures of the DQV instability domains.

To begin with, we note that the hot side of the DBV instability domain
for thin  He-rich envelope models is  virtually the same  than for the
case  of DB  models with  thick  He-rich envelopes.   We have  already
mentioned such  feature in Sect. \ref{template}. This  means that DBVs
with  thick ($M_{\rm He}\sim  10^{-3} M_*$)  or thin  ($M_{\rm He}\sim
10^{-7}-  10^{-8}  M_*$) He-rich  envelopes  should  exhibit the  same
ranges of excited periods, at least in the hot half of the instability
domain.   Both    types   of    DBV   stars   share    the   following
properties.  Firstly, there  is  a strong  dependence  of the  longest
excited periods with  the stellar mass, being larger  for less massive
models. Secondly, the shorter excited periods, on the contrary, do not
exhibit any  dependence with $M_*$.  Finally, the blue edge  is hotter
for more massive DB models.

An interesting feature displayed by  the plots is that the instability
domains of DBVs  and DQVs are clearly separated in  the case of models
with    $M_{\rm   He}=10^{-8}    M_*$   (Figs.     \ref{figure7}   and
\ref{figure9}).   In fact,  there  is a  noticeable  gap in  effective
temperature between the  point at which DB pulsations  no longer exist
and the  location of the  blue edge of  the DQV instability  strip, in
which the pulsations of DQs begin.  This gap is more pronounced in the
case of white dwarfs with $M_*= 0.870\, M_{\odot}$ ($\Delta T_{\rm eff}
\sim 1\,200$ K) than for white dwarfs with $M_*= 0.585\,
M_{\odot}$($\Delta T_{\rm  eff} \sim 400$ K).  For  models with $M_{\rm
He}=2 \times 10^{-7} M_*$, on  the other hand, the instability domains
are  ``in contact''  (Figs.  \ref{figure8}  and  \ref{figure10}). Thus,
there exists a continuous transition from the pulsation instability of
DBs to DQs.  We recall, however, that modes in DBs  are excited by the
maximum  in opacity  due  to  the partial  ionization  of He,  whereas
overstable modes in DQs are mainly driven by the opacity bump produced
by the partial ionization of C.

In Figs.  \ref{figure7} to  \ref{figure10} we have included the periods
detected in the three presently  known DQV stars.  The longest periods
excited  in our  DQ models  are in  the range  $\sim 750-1500$  s. The
periods    detected   in    SDSS   J142625.70$+$575218.4    and   SDSS
J220029.08$-$074121.5 are well within the predicted period ranges, but
the   period  at   $\Pi  \approx   1052$  s   corresponding   to  SDSS
J234843.30$-$094245.3 is too long and is not well accounted for by our
models, except by the case $M_*= 0.585 \, M_{\odot}$ and $M_{\rm He}= 2
\times 10^{-7}  M_*$.  On  the other hand,  only the sequences  of DQs
with  $M_{\rm He}= 10^{-8}  M_*$ are  able to  nearly account  for the
effective temperatures at which the  variable DQ stars are located. As
we  mentioned in  Sect. \ref{theo},  apparently we  could get  rid the
small  discrepancies in  $T_{\rm eff}$  simply by  considering smaller
contents of He in the envelopes.


\section{Summary and conclusions}
\label{conclu}

In this study, we have presented a pulsation stability analysis of the
recently discovered carbon-rich hot DQ white dwarf stars.  These stars
constitute a new class of white dwarf stars, uncovered recently within
the framework  of SDSS  project. There  exist nine hot  DQs, out  of a
total   of   several    thousands   white   dwarfs   spectroscopically
identified. Three  hot DQ white  dwarfs have been reported  to exhibit
photometric   variations  with   periods  compatible   with  pulsation
$g$-modes. One  of the main aims of  the present work was  to test the
convective-mixing  picture  for the  origin  of  hot  DQs through  the
pulsational  properties.   Specifically,  we  have employed  the  full
evolutionary  models of  hot  DQ white  dwarfs  recently developed  by
Althaus et  al.  (2009), that  consistently cover the  whole evolution
from  the born-again  stage  to the  DQ  white dwarf  stage.  We  have
presented  a stability  analysis  on white  dwarf  models from  stages
before  the blue  edge  of  the DBV  instability  strip ($T_{\rm  eff}
\approx  30\,000$ K)  until  the domain  of  the hot  DQ white  dwarfs
($18\,000-24\,000$  K), including  the transition  from DB  to  hot DQ
white  dwarfs.  We  explored  evolutionary models  with $M_*=  0.585\,
M_{\odot}$  and  $M_*=  0.87  \,  M_{\odot}$, and  two  values  for  the
thickness of  the He-rich  envelope ($M_{\rm He}=  2\times 10^{-7}M_*$
and $M_{\rm  He}= 10^{-8} M_*$).   These envelopes are $4-5$  order of
magnitude  thinner  than  those  of  standard DB  white  dwarf  models
resulting from canonical stellar evolution calculations.

We  found  that  at  epochs  when  the  models  are  characterized  by
He-dominated atmospheres, they  exhibit $g$-mode pulsations typical of
DBV  stars,  and when  the  models become  hot  DQ  white dwarfs  with
carbon-dominated   atmospheres,  they  continue   being  pulsationally
unstable with  characteristics similar to  those of DB models,  and in
qualitative agreement with the  periods observationally detected in 
variable hot DQ white dwarfs. Our
main results are the following:

\begin{itemize}

\item Generally, $g$-modes  excited in hot DQ white  dwarfs have small 
linear  growth rates,  noticeably  smaller than  those  typical of  DB
models.  In  spite of this, the  excited modes in hot  DQ white dwarfs
should  have time  enough  as to  reach  observable amplitudes.   This
conclusion is in line with the results of Fontaine et al. (2008).

\item Overstable  $g$-modes  of  DB  white  dwarf  models  having thin 
He-rich envelopes  ($M_{\rm He}/M_*\sim 10^{-8}-10^{-7}$)  are excited
in the same way as in  DB models with thick He-rich envelopes ($M_{\rm
He}/M_*   \sim   10^{-3}$),   that   is,   by  the   action   of   the
$\kappa$-mechanism  acting  on the  region  of  partial ionization  of
He{\sc  ii}.  The opacity  bump due  to the  partial ionization  of C,
present in DB  models with thin envelopes, does  not contribute at all
in the destabilization of modes.

\item Overstable $g$-modes in hot DQ white dwarf  models are primarily 
driven  through  the $\kappa$-mechanism  by  the strong  destabilizing
effect of the  opacity bump due to the partial  ionization of C, being
the  role  of  the partial  ionization  of  He{\sc  ii} of  much  less
relevance. This is somewhat at variance with the models of Fontaine et
al.   (2008),  in which  both  opacity  bumps  are equally  active  in
destabilizing of modes.

\item The blue edge of DQVs is hotter for less massive models than for 
more massive ones. This is at odds with the results of Fontaine et al.
(2008) and Montgomery et al.  (2008).  The reason for this discrepancy
relies in the  way itself in which the DQ  models are conceived, being
in  our case  the result  of  fully evolutionary  computations in  the
context  of  the  diffusive/convective  mixing scenario  presented  by
Althaus et al. (2009).

\item The blue edge for DQV stars is hotter for smaller contents of He 
at the envelopes.

\item The location of the three DQV stars in the $T_{\rm  eff}-\log g$  
diagram is qualitatively accounted for by our calculations of the blue
edge with  $\log(M_{\rm He}/M_*)= -8$, and the  small discrepancies in
$T_{\rm  eff}$ could be  accounted for  by simply  considering stellar
models   with  somewhat   thinner   He-rich  envelopes   ($\log(M_{\rm
He}/M_*)\la -8$).

\item The  prototype star  SDSS  J142625.70$+$575218.4,  that lies  at 
right of our theoretical DQV blue edge, has a surface gravity large in
excess,  and we have  no available  DB/DQ track  that passes  near its
location.  Presumably, this star could have a stellar mass $\ga 1.1 \,
M_{\odot}$.

\item The instability  domains of DBVs and DQVs  are clearly separated 
in  the  $T_{\rm eff}-\Pi$  plane  for the  case  of  models with  the
thinnest   He-rich  envelopes   considered  in   this   work  ($M_{\rm
He}=10^{-8}  M_*$), but  there  is a  continuous  transition from  the
pulsation instability  of DBs  to DQs for  the case in  which $M_{\rm
He}= 2\times 10^{-7} M_*$.

\item The  periods detected  in SDSS J142625.70$+$575218.4   and  SDSS 
J220029.08$-$074121.5  are  well  within  our  theoretical  ranges  of
excited periods, but the period  at $\Pi \approx 1052$ s corresponding
to  SDSS J234843.30$-$094245.3  is long  in  excess, and  is not  well
accounted for by our models.

\end{itemize}

In summary,  our calculations support  the diffusive/convective mixing
picture for the  formation of hot DQs, an  idea originally proposed by
Dufour  et al.  (2008)  and quantitatively  elaborated on  recently by
Althaus et al.  (2009). In particular, the results of the present work
demonstrate that {\sl the diffusive/convective  mixing scenario not only is
able to nicely explain the origin  of hot DQ white dwarfs, but it also
accounts for the variability of these stars.}

According to  this scenario, a fraction  of DB white  dwarfs --- those
which are immediate  progenitors of hot DQ white  dwarfs --- should be
characterized by He-rich envelopes $10^3$ to $10^4$ times thinner than
the  canonical thickness  predicted  by the  standard  theory for  the
formation of DB white dwarfs (Althaus et al.  2005).  The existence of
such DBs,  and thus, the  validity of the  diffusive/convective mixing
scenario, could be investigated using an adiabatic asteroseismological
analysis of DBV  stars with a rich pulsation  spectrum which have high
quality observational data available,  like GD 358, EC 20058-5234, and
CBS 114.

In closing, it is worth  to  stress that the conclusions arrived at in
this  work ---  and also  the results  of Fontaine  et al.  (2008) and
Dufour et al.  (2008) --- in particular concerning the location of the
blue edge of the DBV and DQV instability strips, could  substantially 
change if a fully consistent  treatment of the interaction between convection
and  pulsation like that of Dupret et al. (2008), or even the simpler 
approach of Wu \& Goldreich (1999),  were taken  into  account  in  the 
stability  analysis.


\begin{acknowledgements}

We would like to  warmly acknowledge the comments and suggestions
  by our referee, Dr. M. H.  Montgomery, that allowed us to improve the
  original  version of this  paper.  This  research was  supported by
AGENCIA: Programa de  Modernizaci\'on Tecnol\'ogica BID 1728/OC-AR, by
the  AGAUR, by  MCINN grant  AYA2008--04211--C02--01, by  the European
Union  FEDER funds  and  by PIP  6521  grant from  CONICET.  LGA  also
acknowledges  AGAUR through  the Generalitat  de Catalunya  for  a PIV
grant.   This  research  has  made  use of  NASA's  Astrophysics  Data
System. Finally, we thank H.  Viturro and R.  Mart\'inez for technical
support.

\end{acknowledgements}


\end{document}